\documentclass[%
 aip,
 jcp,%
 amsmath,amssymb,
 reprint,%
final,
]{revtex4-1}
\usepackage{graphicx}
\usepackage{dcolumn}
\usepackage{bm}
\usepackage{color}
\usepackage{xcolor}
\usepackage{tikz}
\providecommand{\e}[1]{\ensuremath{\times 10^{#1}}}
\newcommand{\tm}[1]{\tikz[overlay, remember picture] \node [inner sep=0pt] (#1) { };}
\renewcommand{\vec}[1]{\boldsymbol{\mathbf{#1}}}

\usepackage{calc}
\newlength{\charwidth}
\newlength{\inda}
\newlength{\indaa}
\newlength{\indb}
\newlength{\indbb}
\newlength{\indc}
\newlength{\indcc}
\newlength{\indd}
\newlength{\inddd}
\newcommand{\keyw}[1]{\text{\large{\texttt{#1}}}}

\begin{document}


\title{Stabilized Quasi-Newton Optimization of Noisy Potential Energy
Surfaces} 


\author{Bastian Schaefer}
\affiliation{Department of Physics, University of Basel,
Klingelbergstrasse 82, CH-4056 Basel, Switzerland}

\author{S. Alireza Ghasemi} \affiliation{Institute for Advanced Studies
in Basic Sciences, P.O. Box 45195-1159, IR-Zanjan, Iran}

\author{Shantanu Roy} \affiliation{Computational and Systems Biology,
Biozentrum, University of Basel, CH-4056 Basel, Switzerland}

\author{Stefan Goedecker} \email[]{stefan.goedecker@unibas.ch}
\affiliation{Department of Physics, University of Basel,
Klingelbergstrasse 82, CH-4056 Basel, Switzerland}


\date{\today}

\begin{abstract}
Optimizations of atomic positions belong to the most commonly performed
tasks in electronic structure calculations. Many simulations like
global minimum searches or characterizations of chemical reactions
require performing hundreds or thousands of minimizations or saddle
computations. To automatize these tasks, optimization algorithms must
not only be efficient, but also very reliable. Unfortunately
computational noise in forces and energies is inherent to electronic structure
codes. This computational noise poses a sever problem to the stability
of efficient optimization methods like the limited-memory
Broyden--Fletcher--Goldfarb--Shanno algorithm. We here present a
technique that allows obtaining significant curvature information of
noisy potential energy surfaces. We use this technique to construct
both, a stabilized quasi-Newton minimization method and a stabilized
quasi-Newton saddle finding approach. We demonstrate with the help of
benchmarks that both the minimizer and the saddle finding approach are
superior to comparable existing methods.
\end{abstract}


\maketitle 

\section{Introduction}\label{Introduction}
Stationary points are the most interesting and most important points of
potential energy surfaces. The relative energies of local minima and
their associated configuration space volumes determine thermodynamic
equilibrium properties.\cite{Wales2003} According to transition state
theory, dynamical properties can be deduced from the energies and the
connectivity of minima and transition states.\cite{Eyring1935}
Therefore, the efficient determination of stationary points of
potential energy surfaces is of great interest to the communities of
computational chemistry, physics, and biology.  Clearly, optimization
and in particular minimization problems are present in virtually any
field.  This explains why the development and mathematical
characterization of iterative optimization techniques are important and
longstanding research topics, which resulted in a number of highly
sophisticated methods like for example direct inversion of the
iterative subspace (DIIS),\cite{Pulay1980,Pulay1982}
conjugate gradient (CG),\cite{Stiefel1952} or quasi-Newton methods like the
Broyden-Fletcher-Goldfarb-Shanno (BFGS) algorithm\cite{Broyden1970,
Fletcher1970,Goldfarb1970,Shanno1970} and its limited memory variant
(L-BFGS).\cite{Nocedal1980,Liu1989} Since for a quadratic function
Newton's method is guaranteed to converge within a single iteration, it
is not surprising that the BFGS and L-BFGS algorithms belong to the
most efficient methods for minimizations of atomic
systems.\cite{Wales2003}

If the potential energy surface can be computed with an accuracy on the
order of the machine precision, the above mentioned algorithms usually
work extremely well. In practice, however, computing the energy surface at
this high precision is not possible for physically accurate but
computationally demanding levels of theory like for example density
functional theory (DFT). At DFT
level, this is due the finitely spaced integration grids and self
consistency cycles that have to be stopped at small, but non-vanishing
thresholds.  Therefore, optimization
algorithms that are used at these accurate levels of theory must
not only be computationally efficient but also tolerant to noise in
forces and energies.
Unfortunately, the very efficient L-BFGS algorithm is known to be
noise-sensitive and therefore, frequently fails to converge on noisy
potential energy surfaces. For this
reason, the fast inertial relaxation engine (FIRE) has been
developed.\cite{Bitzek2006} FIRE is a method of the damped molecular
dynamics (MD) class of optimizers.\cite{Tassone1994,Probert2003} It accelerates convergence by mixing
the velocity at every MD step with a fraction of the current steepest
descent direction. A great advantage of FIRE is its simplicity. However,
FIRE does not make use of any curvature information and therefore
usually is significantly less efficient than the Newton or quasi-Newton methods.

Potential energy surfaces are bounded from below and therefore descent
directions guarantee that a local minimum will finally be found.
Furthermore, the curvature at a minimum is positive in all
directions. This means, all directions can be treated on the same
footing during a minimization.  The situation is different for saddle
point optimizations.  A saddle point is a stationary point at which
the potential energy surface is at a maximum with respect to
one or more particular directions and at a minimum with respect to  all
other directions. Close to a saddle point it is therefore not
possible to treat all directions on the same footing. Instead one has to
single out the directions that have to be maximized. Furthermore, far
away from a saddle point it is usually impossible to tell, which
search direction guarantees to finally end up in a saddle point.
Therefore, saddle point optimizations typically are
more demanding and significantly less reliable than minimizations.

In this contribution we present a technique that allows to extract
curvature information from noisy potential energy surfaces. We explain
how to use this technique to construct a stabilized quasi-Newton
minimizer (SQNM) and a stabilized quasi-Newton saddle finding method
(SQNS).  Using benchmarks, we demonstrate that both optimizers are
robust and efficient. The comparison of SQNM to L-BFGS and FIRE and of
SQNS to an improved dimer method\cite{Henkelman1999,Kaestner2008} reveals that SQNM and SQNS are
superior to their existing alternatives.

\section{Methods}
\subsection{Newton's and Quasi Newton's Method}\label{sec:newtonsmethod}
The potential energy surface of an $N$-atomic system is a map
$E:\mathbb{R}^{3N}\mapsto \mathbb{R}$ that assigns to each atomic
configuration $\vec{R}$ a potential energy. It is assumed that a second
order expansion of $E\left(\vec{R}\right)$ about a point
$\vec{R}^{i}$ is possible:
\begin{align}
E\left(\vec{R}\right) &\approx E\left(\vec{R}^{i}\right)+\left[\vec{R}-
                         \vec{R}^{i}\right]^{T}\vec{\nabla}
                         E\left(\vec{R}^{i}\right)\notag\\
&\phantom{{}=E\left(\vec{R}^{i}\right)}+ \frac{1}{2} \left[\vec{R}-
             \vec{R}^{i}\right]^{T} H_{\vec{R}_{i}} \left[\vec{R}-
             \vec{R}^i\right]\\
\vec{\nabla}E\left(\vec{R}\right) &\approx \vec{\nabla}
           E\left(\vec{R}^{i}\right) + H_{\vec{R}_{i}}\left[\vec{R}-
           \vec{R}^i\right], \label{eq:secant}
\end{align}
Here, $H_{\vec{R}^{i}}$ is the Hessian of the potential energy surface
evaluated at $\vec{R}^{i}$. If $\vec{R}$ is a stationary point, the
left hand side gradient of Eq.~\ref{eq:secant} vanishes and Newton's
optimization method follows:
\begin{align}
\vec{R}^{i+1} &= \vec{R}^{i}-H^{-1}_{\vec{R}_{i}}
                 \vec{\nabla}E\left(\vec{R}^{i}\right)\label{eq:newton}
\end{align}
In the previous equation $\vec{R}$ was renamed to $\vec{R}^{i+1}$ in
order to emphasize the iterative character of Newton's Method for
non-quadratic potential energy surfaces.

In practice, it is in most cases either impossible to calculate an analytic
Hessian or it is too time consuming to compute it numerically
by means of finite differences at every iteration. Therefore,
quasi-Newton methods use an approximation to the exact Hessian that is
computationally less demanding. Using a constant multiple of the
identity matrix as an approximation to the Hessian results in the
simple steepest descent method. In most cases, such a choice is a very
poor approximation to the true Hessian. However, improved
approximations can be generated from local curvature information which
is obtained from the history of the last
$n_{\text{hist}}$
displacements $\vec{\Delta{R}}^{i}
:=\vec{R}^{i}-\vec{R}^{i-1}$ and gradient differences
$\vec{\Delta{g}}^{i} := \vec{\nabla} E\left(R^{i}\right)-\vec{\nabla}
E\left(\vec{R}^{i-1}\right)$, where $i=1\dots n_{\text{hist}}$.

\subsection{Significant Subspace in Noisy Optimization
Problems}\label{sec:sigsub}
In noisy optimization problems, the noisy components of the
gradients can lead to displacement components that correspond to
erratic movements on the potential energy surface.
Consequently, curvature information that comes from the subspace
spanned by these displacement components must not be used for the
construction of an approximate Hessian. In contrast to this, the
non-noisy gradient components promote locally systematic net-movements,
which do not tend to cancel each other.  In this sense the displacement
components that correspond to these well defined net-movement span a
significant subspace from which meaningful curvature information can be
extracted and used for building an approximate Hessian.

\begin{figure}
    \centering
    \includegraphics{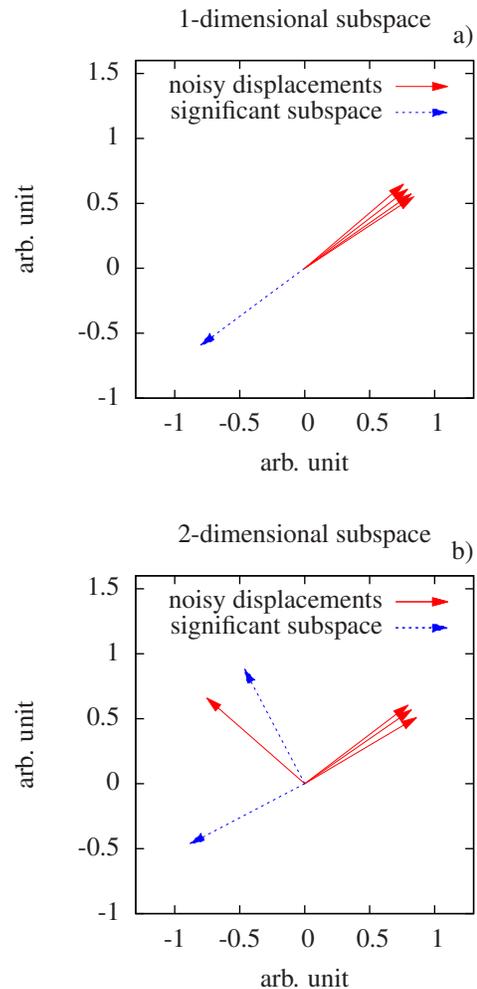}
    \caption{Illustrated are significant subspaces spanned by the
displacements in a model atomic coordinate space. Only from the significant subspace
it is meaningful to extract curvature information. The red solid
arrows simulate displacements made under the influence of noisy forces.
The blue dashed arrows show significant subspaces from which it is
meaningful to extract curvature information. Panel a) shows a case in
which the significant subspace is only one-dimensional. Panel b) shows
an example in which curvature information can be extracted from the
full 2-dimensional space.  The significant subspaces that are shown
here were computed using the method outlined in section~\ref{sec:sigsub}}
    \label{fig:removenoise}
\end{figure}
The situation is depicted in Fig.~\ref{fig:removenoise} where the red
solid vectors represent the history of normalized
displacements and the blue dashed vectors constitute a basis of the
significant subspace.  All the red solid vectors in
Fig.~\ref{fig:removenoise}a point into similar directions. Therefore,
curvature information should only be extracted from a one-dimensional
subspace, as, for example, is given by the blue dashed vector.
Displacement components  perpendicular to this blue dashed vector come
from the noise in the gradients.
In contrast to Fig.~\ref{fig:removenoise}a, Fig.~\ref{fig:removenoise}b shows a displacement
that points into a considerably different direction than all the other
displacements.  For this reason, significant curvature information
can be extracted in the full two-dimensional space.

To define the significant subspace more rigorously,
we first introduce the set of normalized displacements
\begin{align}
\widehat{\vec{\Delta{R}}}^{i}&:=\frac{\vec{\Delta{R}}^{i}}{|\vec{\Delta{R}}^{i}|},
\end{align}
where $i=1\dots n_{\text{hist}}$.
With $\sum_k |\vec{\omega}_{k}|^2=1$, linear combinations $\vec{w}$ of the normalized
displacements are
defined as:
\begin{align}
\vec{w}:= \sum_{k=1}^{n_{\text{hist}}}
\vec{\omega}_{k}\widehat{\vec{\Delta{R}}}^{k}\label{eqn:lincomb},
\end{align}
Furthermore, we define a real symmetric overlap matrix
$S$ as
\begin{align}
S_{kl} &:=\widehat{\vec{\Delta{R}}}^{k}\cdot
\widehat{\vec{\Delta{R}}}^{l}.
\end{align}
It can be seen from,
\begin{align}
\vec{w}\cdot\vec{w}&=\vec{\omega}^T S \vec{\omega},
\end{align} that $|\vec{w}|$ is made stationary by coefficient vectors
$\vec{\omega}^i$ that are eigenvectors of the overlap matrix.  In
particular the longest and shortest vectors that can be generated by
linear combinations with normalized coefficient vectors $\vec{\omega}$
correspond to those eigenvectors of the overlap matrix that have the largest
and smallest eigenvalues.  As motivated above, the shortest linear
combinations of the normalized displacements correspond to noise.

From now on, let the $\vec{\omega}^{i}$ be eigenvectors of
$\left(S_{kl}\right)$ and let $\lambda_i$ be the corresponding
eigenvalues.
With
\begin{align}
\stackrel{\sim}{\smash{\widehat{\vec{\Delta{R}}}^{i}}\rule{0pt}{2.1ex}}&:=
\frac{1}{\sqrt{\lambda_i}}\sum_{k=1}^{n_{\text{hist}}}
\vec{\omega}^{i}_{k}\widehat{\vec{\Delta{R}}}^{k},
\end{align}
we finally define the \textit{significant subspace} $\mathfrak{S}$ as
\begin{align}
\mathfrak{S}&:=\text{span}\left(\left\{\stackrel{\sim}{\smash{\widehat{\vec{\Delta{R}}}^{i}}\rule{0pt}{2.1ex}}
\;\middle|\; \lambda_i/\underset{j}{\mathrm{max}}\left\{\lambda_{j}\right\}
> \epsilon\right\}\right),
\end{align}
where $0\le\epsilon\le1$.  In all
applications presented in this work, $\epsilon=10^{-4}$ has proven to
work well.  Henceforth, we will refer to the dimension of
$\mathfrak{S}$ as $n_{\text{dim}}$. By construction it is guaranteed
that $n_{\text{dim}} \le 3N$.  It should be noted that at each
iteration of the optimization algorithms that are introduced
below, the significant subspace and its dimension
$n_{\text{dim}}$ can change. The history length
$n_{\text{hist}}$ usually lies between 5 and 20.

Our procedure is analogous to L\"{o}wdins canonical
orthogonalization,\cite{Loewdin1956, Mayer2003, Jensen2007} which is
used in the electronic structure community to remove linear
dependencies from chemical basis sets.

\subsection{Obtaining Curvature Information on the Significant Subspace}
We define the projection $\overset{\sim}{H}$ of the Hessian $H$ onto
$\mathfrak{S}$ as
\begin{align}
\overset{\sim}{H} &:= PHP\notag\\
&=\sum_{ij}H_{ij}\stackrel{\sim}{\smash{\widehat{\vec{\Delta{R}}}^{i}}\rule{0pt}{2.1ex}}
\left(\stackrel{\sim}{\smash{\widehat{\vec{\Delta{R}}}^{j}}\rule{0pt}{2.1ex}}\right)^{T},
\end{align}
where for all $
\stackrel{\sim}{\smash{\widehat{\vec{\Delta{R}}}^{i}}\rule{0pt}{2.1ex}}\in
\mathfrak{S}$
$$P :=
\sum_{i=1}^{n_{\text{dim}}}\stackrel{\sim}{\smash{\widehat{\vec{\Delta{R}}}^{i}}\rule{0pt}{2.1ex}}
\left(\stackrel{\sim}{\smash{\widehat{\vec{\Delta{R}}}^{i}}\rule{0pt}{2.1ex}}\right)^{T}$$
and
$$H_{ij} :=
\left(\stackrel{\sim}{\smash{\widehat{\vec{\Delta{R}}}^{i}}\rule{0pt}{2.1ex}}\right)^{T}
H
\stackrel{\sim}{\smash{\widehat{\vec{\Delta{R}}}^{j}}\rule{0pt}{2.1ex}}.$$
Using Eq.~\ref{eq:secant} and defining
\begin{align}
\stackrel{\sim}{\smash{\vec{\Delta{g}}^{i}}\rule{0pt}{1.3ex}} &:=
\frac{1}{\sqrt{\lambda_i}}\sum_{k=1}^{n_{\text{hist}}}
\frac{\vec{\omega}^{i}_{k}}{|\Delta \vec{R}^{k}|}\Delta
\vec{g}^k,\label{eq:deltag}
\end{align}
where $i=1\dots n_{\text{dim}}$,
one obtains an approximation for each matrix element $H_{ij}$:
\begin{align}
{H}_{ij} &\approx
\stackrel{\sim}{\smash{\vec{\Delta{g}}^{i}}\rule{0pt}{1.3ex}}
\cdot
\stackrel{\sim}{\smash{\widehat{\vec{\Delta{R}}}^{j}}\rule{0pt}{2.1ex}}.
\end{align}
In practice, we explicitly symmetrize $H_{ij}$ in order to avoid
asymmetries introduced by anharmonic effects:
\begin{align}
H_{ij} &\approx
\frac{1}{2}
\left(
\stackrel{\sim}{\smash{\vec{\Delta{g}}^{i}}\rule{0pt}{1.3ex}}
\cdot
\stackrel{\sim}{\smash{\widehat{\vec{\Delta{R}}}^{j}}\rule{0pt}{2.1ex}}
+
\stackrel{\sim}{\smash{\vec{\Delta{g}}^{j}}\rule{0pt}{1.3ex}}
\cdot
\stackrel{\sim}{\smash{\widehat{\vec{\Delta{R}}}^{i}}\rule{0pt}{2.1ex}}
\right).
\end{align}
Because the projection $P$ is the identity operator on $\mathfrak{S}$,
the curvature $c(\widehat{\vec{d}})$ on the potential energy surface
along a normalized $\widehat{\vec{d}}\in \mathfrak{S}$ is given by
\begin{align}
c(\widehat{\vec{d}}) &= \widehat{\vec{d}}^T
\overset{\sim}{H}\widehat{\vec{d}}.\label{eq:curv}
\end{align}
Given the normalized eigenvectors $\vec{v}^{i}$ and corresponding
eigenvalues $\kappa_i$ of the $n_{\text{dim}}\times n_{\text{dim}}$
Matrix        $\left(H_{ij}\right)$, one can write normalized
eigenvectors $\overset{\sim}{\vec{v}}_{i}\in\mathfrak{S}$ of
$\overset{\sim}{H}$ with eigenvalues $\kappa_i$ as
\begin{align}
\overset{\sim}{\vec{v}}^{i} =
\sum_{k=1}^{n_{\text{dim}}}\vec{v}^{i}_{k}\stackrel{\sim}{\smash{\widehat{\vec{\Delta{R}}}^{k}}\rule{0pt}{2.1ex}},\label{eq:eigenvec}
\end{align}
where $\vec{v}^{i}_{k}$ is the k-th element of $\vec{v}^{i}$.
As can be seen from Eq.~\ref{eq:curv}, the $\kappa_i$ give the curvatures
of the potential energy surface along the directions
$\overset{\sim}{\vec{v}}^{i}$.

\subsection{Using Curvature Information on the Significant Subspace for
Preconditioning
$\vec{\nabla}E$}\label{sec:precondSigGrad}
The gradient $\vec{\nabla} E$ can be decomposed into a component lying
in $\mathfrak{S}$ and a component lying in its orthogonal complement:
\begin{align} \vec{\nabla} E &= \vec{\nabla} E_{\mathfrak{S}} +
\vec{\nabla}E_{\perp}, \end{align} where $\vec{\nabla}
E_{\mathfrak{S}}:=P'\vec{\nabla} E $,
$\vec{\nabla}E_{\perp}:=(I-P')\vec{\nabla} E\notag$ and $P':=\sum_{i}
\overset{\sim}{\vec{v}}^{i}\left(\overset{\sim}{\vec{v}}^{i}\right)^T$.
In this section we motivate how the $\kappa_i$ can be used to
precondition $\vec{\nabla}E_{\mathfrak{S}}$. Furthermore, we explain
how $\vec{\nabla}E_{\perp}$ can be scaled appropriately with the help
of a feedback that is based on the angle between two consecutive
gradients.

Let us assume that the Hessian $H$ at the current point of the potential energy
surface is non-singular and let $\nu_i$ and $\vec{V}^i$ be its
eigenvalues and normalized eigenvectors.  In Newton's Method
(Eq.~\ref{eq:newton}), the gradients are conditioned by the inverse
Hessian. For the significant subspace component
$\vec{\nabla}E_{\mathfrak{S}}$ it follows:
\begin{align}
H^{-1}\vec{\nabla}E_{\mathfrak{S}}&=\sum_{i=1}^{3N}\sum_{j=1}^{n_{\text{dim}}}\left[\left(\frac{\vec{\nabla}E\cdot\overset{\sim}{\vec{v}}^{j}}{\nu_i}\right)\left(\overset{\sim}{\vec{v}}^{j}\cdot\vec{V}^i\right)\vec{V}^{i}\right]\label{eq:firstidea}
\end{align}
As outlined in the previous section, we know the curvature $\kappa_j$
along $\overset{\sim}{\vec{v}}^{j}$. Therefore, at a first thought,
Eq.~\ref{eq:firstidea} suggests to simply replace $\nu_i$ by $\kappa_j$
where $i=1\dots 3N$ and $j=1\dots n_{\text{dim}}$.
Indeed, if the optimization was restricted to the subspace
$\mathfrak{S}$ this choice would be appropriate. However, with respect
to the complete domain of the potential energy surface, one is at risk
to underestimate the curvature $\nu_i$ if the overlap
$O_{ij}:=\overset{\sim}{\vec{v}}^{j}\cdot\vec{V}^i$ is  non-vanishing.

In particular, if $O_{ij}$ is far from being negligible,
underestimating the curvature $\nu_i$ can be particularly problematic
because coordinate changes in the direction of $\vec{V}^i$ might be too
large. This can render convergence difficult to obtain in practice.

We therefore replace $\nu_i$ in Eq.~\ref{eq:firstidea} by
\begin{align}
\kappa_j':=\sqrt{\kappa_j^2 + r_j^2},
\end{align}
where $r_j$ is chosen in analogy
to the residue of Weinstein's Criterion\cite{Weinstein1934,Suzuki1998}
as
\begin{align}
r_j &:=
\left|H\overset{\sim}{\vec{v}}^{j} -
\left((\overset{\sim}{\vec{v}}^{j})^TH\overset{\sim}{\vec{v}}^{j}\right)\overset{\sim}{\vec{v}}^{j}\right|.
\end{align}
Using equations~\ref{eq:deltag},~\ref{eq:curv} and~\ref{eq:eigenvec},
this residue can be approximated by
\begin{align}
r_j &\approx
\left|\sum_{k=1}^{n_{\text{dim}}}\left[\vec{v}_{k}^{i}
\stackrel{\sim}{\smash{\vec{\Delta{g}}^{k}}\rule{0pt}{1.3ex}}\right] -
\kappa_{j} \overset{\sim}{\vec{v}}^{j}\right|.
\end{align}
With this choice for $\kappa_j'$, the preconditioned gradient
$\vec{\nabla}E_{\mathfrak{S}}^{\text{P}}$ is finally given by:
\begin{align}
\vec{\nabla}E_{\mathfrak{S}}^{\text{P}}&:=
\sum_{j=1}^{n_{\text{dim}}}\left(\frac{\vec{\nabla}E\cdot\overset{\sim}{\vec{v}}^{j}}{\kappa_j'}\right)\overset{\sim}{\vec{v}}^{j}.
\end{align}
\begin{figure}
    \centering
    \includegraphics{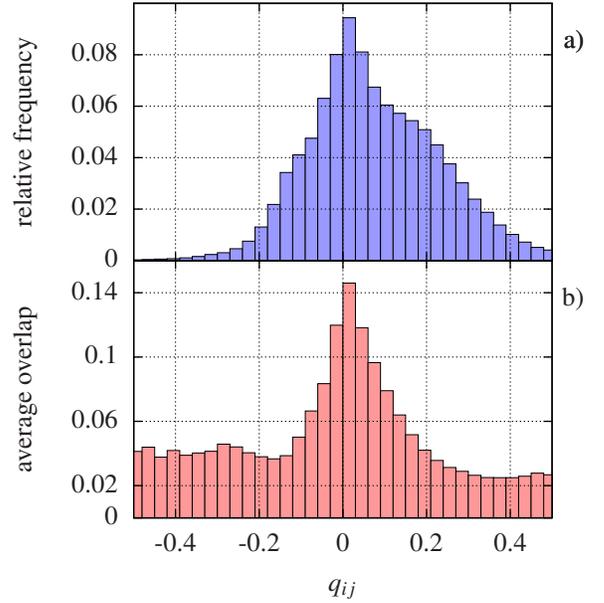}
    \caption{Panel a) is a histogram of
$q_{ij}:=\sqrt{\kappa_{j}^{2}+r_{j}^{2}}-\nu_i$ for $i=1\dots3N$ and
$j=1\dots n_{\text{dim}}$. $q_{ij}$ is a measure for the quality of the
estimation of the eigenvalue $\nu_i$ of the exact Hessian.  Panel b)
shows the bin-averaged overlap $O_{ij}$.  The frequency of severe
curvature underestimation drops quickly in the region $q_{ij} < 0$. The
histogram in panel a) peaks in the region of good estimation ($q_{ij} \approx
0$) which coincidences with the region of large overlap $O_{ij}$, shown in
panel b).  The data for this figure come from 100 minimizations of a
$\text{Si}_{20}$ system described by the
Lenosky-Silicon\cite{Lenosky2000,Goedecker2002} force field.}
    \label{fig:weinstein}
\end{figure}
Clearly, the residue $r_j$ can only alleviate the problem of curvature
underestimation, but it does not rigorously guarantee that every single
$\nu_i$ is estimated appropriately. However, in practice this choice
works very well. The reason for this can be seen from
Fig.~\ref{fig:weinstein}. In Fig.~\ref{fig:weinstein}a, a
histogram of the quality and safety measure
$q_{ij}:=\sqrt{\kappa_{j}^{2}+r_{j}^{2}}-\nu_i$ is shown. If $q_{ij}
<0$, the curvature $\nu_i$ is underestimated, if $q_{ij}\approx 0$ the
curvature $\nu_i$ is well estimated and finally, if $q_{ij}>0$, the
curvature is overestimated. Overestimation leads to too small
step sizes, and therefore to a more stable algorithm, albeit at the
cost of a performance loss. Critical underestimation of the curvature
($q_{ij} \ll 0$) is rare. Fig.~\ref{fig:weinstein}b shows the averages of the overlap
$O_{ij}$ in the corresponding bins.  If $\overset{\sim}{\vec{v}}^{j}$
has on average a large overlap with $\vec{V}^i$, the curvature along
$\vec{V}^i$ is estimated accurately (histogram in Fig.~\ref{fig:weinstein}a peaks at $q_{ij} \approx 0$).

What remains to discuss is how the gradient component
$\vec{\nabla}E_{\perp}$ should be scaled. By construction,
$\vec{\nabla}E_{\perp}$ lies in the subspace for which no curvature
information is available. We therefore treat this gradient component by
a simple steepest descent approach that adjusts the step size $\alpha
\in \mathbb{R}^{+}$
at each iteration. For the minimizer that is outlined in
section~\ref{sec:findmin}, the adjustment is based on the angle between
the complete gradient $\vec{\nabla}E$ and the preconditioned gradient $\vec{\nabla}E^{\text{P}}$. If the cosine of this
intermediate angle is larger than $0.2$, $\alpha$ is increased by a factor of
$1.1$, otherwise $\alpha$ is decreased by a factor of $0.85$.
For the saddle search algorithm the feedback is slightly different and
will be explained in section~\ref{sec:findsad}.

In conclusion, the total preconditioned gradient $\vec{\nabla}E^{\text{P}}$ is given by
\begin{align}
\vec{\nabla}E^{\text{P}} &:= \vec{\nabla}E_{\mathfrak{S}}^{\text{P}} + \alpha \vec{\nabla}E_{\perp}
\end{align}
In the next section, we explain how this preconditioned gradient can be
further improved for biomolecules.

The preconditioned subspace gradient
$\vec{\nabla}E_{\mathfrak{S}}^{\text{P}}$ was obtained under the
assumption of a quadratic potential energy surface. However, if the
gradients at the current iteration are large, this assumption is probably
not satisfied. Displacing along
$\vec{\nabla}E_{\mathfrak{S}}^{\text{P}}$ in these cases can reduce the stability of
the optimization. Hence, if the $|\vec{\nabla}E|$ exceeds a
certain threshold, it can be useful  to set the dimension of
$\mathfrak{S}$ to zero for a certain number of iterations. This means
that
$\vec{\nabla}E_{\perp} = \vec{\nabla}E$ and therefore
$\vec{\nabla}E^{\text{P}}=  \alpha \vec{\nabla}E $. In that case,
$\alpha$ is also adjusted according to the above described gradient
feedback. However, as this fallback to steepest descent is intended
as a last final fallback, it should have the ability to deal with
arbitrarily large forces. Therefore, we also check that $\alpha
\vec{\nabla}E$ does not displace any atom by more than a user-defined
trust radius. However, to our experience, this fallback
is not necessary in most cases. Indeed, all the benchmarks
presented in section~\ref{sec:bench} were performed without this
fallback.

\subsection{Additional Efficiency for Biomolecules}\label{sec:bioprec}
Many large molecules like biomolecules or polymers are floppy systems
in which the largest and
smallest curvatures can be very different from each other. Steepest
descent optimizers are very inefficient for these ill-conditioned
systems, because the high curvature directions force to use step sizes
that are far too small for an efficient optimization in the directions
of small curvatures. Put more formally, the optimization is
inefficient for those systems, because the condition number, which is
the fraction of largest and smallest curvature, is large.\cite{Goedecker2001}
For biomolecules, the high-curvature directions usually correspond to bond stretchings,
that is, movements along inter-atomic displacement vectors of bonded
atoms. For the current purpose we regard two atoms to be bonded if
their inter-atomic distance is smaller than or equal to $1.2$ times the
sum of their covalent radii. For $i=1\dots N$, let $\vec{r}^{i}\in
\mathbb{R}^3$ be the coordinate vector  of the i-th atom. For a system
with $n_{\text{bond}}$ bonds we define for each bond a bond vector
$\vec{b}^{m} \in \mathbb{R}^{3N}$, $m=1\dots n_{\text{bond}}$
\begin{align}
\vec{b}^{m} &:= \left( \begin{matrix} \overset{\sim}{\vec{b}}^{m,1} \\
\overset{\sim}{\vec{b}}^{m,2} \\ \vdots \\ \overset{\sim}{\vec{b}}^{m,N}
\end{matrix} \right),
\end{align}
where the $\overset{\sim}{\vec{b}}^{m,k} \in \mathbb{R}^3$, $k=1\dots N$
are defined as
\begin{alignat}{2}
\overset{\sim}{\vec{b}}^{m,i} &:= -\overset{\sim}{\vec{b}}^{m,j} & &:=\begin{cases}
\vec{r}^{j}-\vec{r}^{i}, & \text{if atoms i and j are}\\
                         & \text{bonded by the m-th
bond}.\\
    \vec{0}, & \text{otherwise}.
  \end{cases}
\end{alignat}
The $\vec{b}^m$ are sparse vectors with six non-zero elements.

We separate the total gradient $\vec{\nabla}E$ into its
bond-stretching components $\vec{\nabla}E_{\text{str}}$ and all the
remaining components $\vec{\nabla}E_{\text{r}}$:
\begin{align}
\vec{\nabla}E &= \vec{\nabla}E_{\text{str}} +
\vec{\nabla}E_{\text{r}}.\label{eq:stretchsep}
\end{align}
Let $c_m \in \mathbb{R}$ be coefficients that allow the bond-stretching
components to be expanded in terms of the bond vectors
\begin{align}
\vec{\nabla}E_{\text{str}} &:= \sum_{m=1}^{n_{\text{bond}}}
c_{m}\label{eq:stretchgrad}
\vec{b}^{m}.
\end{align}
Using definition Eq.~\ref{eq:stretchgrad}, left-multiplying
Eq.~\ref{eq:stretchsep} with a bond vector $\vec{b}^{n}$ and requiring
the $\vec{\nabla}E_{\text{r}}$ to be orthogonal to all the bond
vectors, one obtains the following linear system of equations, which
determines the coefficients $c_m$ and, with it, the bond stretching
gradient defined in Eq.~\ref{eq:stretchgrad}:
\begin{align}
\vec{b}^{n}\cdot \vec{\nabla}E &= \sum_{m} c_m \vec{b}^{n} \cdot
\vec{b}^{m}.
\end{align}

For the optimization of a biomolecule, the bond-stretching components
are minimized in a simple steepest descent fashion. The atoms are
displaced by $-\alpha_{\text{s}}\vec{\nabla}E_{\text{str}}$. The
bond-stretching step size $\alpha_{\text{s}}$ is a positive constant,
which is adjusted in each iteration of the optimization by simply
counting the number of projections $\vec{b}^{m}\cdot \vec{\nabla}E$
that have not changed signs since the last iteration. If more than two
thirds of  the signs of the projections have remained unchanged, the
bond-stretching step size $\alpha_{\text{s}}$ is increased by 10
percent. Otherwise, $\alpha_{\text{s}}$ is decreased by a factor of
$1/1.1$. The non-bond-stretching gradients $\vec{\nabla}E_{\text{r}}$
are preconditioned using the stabilized quasi-Newton approach presented
in sections~\ref{sec:sigsub} to~\ref{sec:precondSigGrad}. It is
important to note that in sections~\ref{sec:sigsub}
to~\ref{sec:precondSigGrad} all $\vec{\nabla}E$ have to be replaced by
$\vec{\nabla}E_{\text{r}}$ when using this biomolecule preconditioner.
In particular, this is also true for the gradient feedbacks that are
described in sections~\ref{sec:precondSigGrad} and~\ref{sec:findsad}.

\subsection{Finding Minima -- The SQNM method}\label{sec:findmin}
The pseudo code below demonstrates how the above presented techniques
can be assembled into an efficient and stabilized quasi-newton
minimizer (SQNM). The pseudo code contains 4 parameters explicitly.
$\alpha_{\text{start}}$ and $\alpha_{\text{s,start}}$ are initial step
sizes that scale $\vec{\nabla} E_{\perp}$ and $\vec{\nabla}
E_{\text{str}}$, respectively. $m$ is the maximum length of the history
list from which the significant subspace $\mathfrak{S}$ is constructed.
$E_{\text{thresh}}$ is an energy-threshold that is used to determine
whether a minimization step is accepted or not. It should be adapted to
the noise level of the energies and forces.  The history list is
discarded if the energy increases, because an increase in energy is an
indication for inaccurate curvature information. In this case, the
dimension of the significant subspace is considered to be zero.
Furthermore, line 17 implicitly contains the parameter $\epsilon$,
which is described in section~\ref{sec:sigsub}.  The optimization is
considered to be converged if the norm of the gradient is smaller than
a certain threshold value. Of course, other force criteria, like for
example using the maximum force component instead of the force norm,
are possible.
\begingroup
\allowdisplaybreaks
\begin{flalign*}
\text{ 1. }  &\alpha \leftarrow \alpha_{\text{start}};\text{ }
             \alpha_s \leftarrow \alpha_{\text{s,start}}; &\\
\text{ 2. }  &\text{accepted} \leftarrow \text{ true}; &\\
\text{ 3. }  &k \leftarrow 1; &\\
\text{ 4. }  &\text{Initialize }\vec{R}_k\text{ with coordinates;} &\\
\text{ 5. }  &E_k \leftarrow E(\vec{R}_k); &\\
\text{ 6. }  &\keyw{repeat} &\\
\text{ 7. }  &\hspace{\inda}\keyw{if}  \text{ optimizing biomolecule } \keyw{then} &\\
\text{ 8. }  &\hspace{\indb}\keyw{if }\text{accepted }\keyw{then}  &\\
\text{ 9. }  &\hspace{\indc}\parbox[t]{\dimexpr\linewidth-\indcc}{
              \setlength{\hangindent}{0.5\inda}Compute
              $\vec{\nabla}E_{\text{str}}$ for $\vec{R}_{k}$, as
              outlined in section~\ref{sec:bioprec};} &&\\
\text{10. }  &\hspace{\indc}\parbox[t]{\dimexpr\linewidth-\indcc}{
              \setlength{\hangindent}{0.5\inda}Adjust $\alpha_s$ based
              on the feedback described in section~\ref{sec:bioprec};} &&\\
\text{11. }  &\hspace{\indc}\vec{g}_k \leftarrow \vec{\nabla}E(\vec{R}_{k}) - \vec{\nabla}E_{\text{str}}; &\\
\text{12. }  &\hspace{\indc} \vec{R}_{k} \leftarrow \vec{R}_{k}  - \alpha_s \vec{\nabla}E_{\text{str}};&\\
\text{13. }  &\hspace{\indb}\keyw{end if }  &\\
\text{14. }  &\hspace{\inda}\keyw{else}   &\\
\text{15. }  &\hspace{\indb}\vec{g}_k \leftarrow \vec{\nabla}E(\vec{R}_{k}); &\\
\text{16. }  &\hspace{\inda}\keyw{end if }  &\\
\text{17. }  &\hspace{\inda}\parbox[t]{\dimexpr\linewidth-\indaa}{
              \setlength{\hangindent}{0.5\inda}
              Based on the $\{\vec{g}_j,\vec{R}_j\}_{j\le k}$ in the
history list, compute the preconditioned
              gradient $\vec{\nabla}E^{\text{P}}$ as outlined
              in sections~\ref{sec:sigsub}~to~\ref{sec:bioprec};}&\\
\text{18. }  &\hspace{\inda} \vec{R}_{k+1} \leftarrow \vec{R}_{k} -
              \vec{\nabla}E^{\text{P}};&\\
\text{19. }  &\hspace{\inda}\parbox[t]{\dimexpr\linewidth-\indaa}{\setlength{\hangindent}{0.5\inda}
              \keyw{if } $E(\vec{R}_{k+1}) > E_{k} + E_{\text{thresh}}$
               \keyw{ and } \mbox{$\alpha>\alpha_{\text{start}}/10$} \keyw{then}}&\\
\text{20. }  &\hspace{\indb}\parbox[t]{\dimexpr\linewidth-\indbb}{
              \setlength{\hangindent}{0.5\inda} accepted $\leftarrow$ false;}&\\
\text{21. }  &\hspace{\indb}\parbox[t]{\dimexpr\linewidth-\indbb}{
              \setlength{\hangindent}{0.5\inda} Remove
              $\{\vec{g}_{j},\vec{R}_{j}\}_{j<k}$ from the history list;}&\\
\text{22. }  &\hspace{\indb}\parbox[t]{\dimexpr\linewidth-\indbb}{
              \setlength{\hangindent}{0.5\inda}
              $\alpha \leftarrow \alpha/2$;}&\\
\text{23. }  &\hspace{\inda}\keyw{else } &\\
\text{24. }  &\hspace{\indb}\parbox[t]{\dimexpr\linewidth-\indbb}{
              \setlength{\hangindent}{0.5\inda} accepted $\leftarrow$ true;}&\\
\text{25. }  &\hspace{\indb}\parbox[t]{\dimexpr\linewidth-\indbb}{
              \setlength{\hangindent}{0.5\inda} $E_{k+1} \leftarrow E(\vec{R}_{k+1})$;}&\\
\text{26. }  &\hspace{\indb}\parbox[t]{\dimexpr\linewidth-\indbb}{
              \setlength{\hangindent}{0.5\inda}Adjust $\alpha$ based on
              the gradient feedback described in section~\ref{sec:precondSigGrad};}&\\
\text{27. }  &\hspace{\indb}\keyw{if } k > m \keyw{ then} &\\
\text{28. }  &\hspace{\indc}\parbox[t]{\dimexpr\linewidth-\indcc}{
              \setlength{\hangindent}{0.5\inda} Remove
              $\vec{R}_{k-m}$ and $\vec{g}_{k-m}$ from
              storage;}&\\
\text{29. }  &\hspace{\indb}\keyw{end if }&\\
\text{30. }  &\hspace{\indb}k \leftarrow k + 1; &\\
\text{31. }  &\hspace{\inda}\keyw{end if }&\\
\text{32. }  &\keyw{until convergence.} &\\
\end{flalign*}%
\endgroup

\subsection{Finding Saddle Points -- The SQNS Method}\label{sec:findsad}
In this section we describe a stabilized quasi-Newton saddle finding method
(SQNS) that is based on the same principles as the minimizer in the
previous section. SQNS belongs to the class of the minimum mode
following methods.\cite{Cerjan1981,Wales1993,Henkelman1999}

For simplicity, we will denote the Hessian eigenvector corresponding to
the smallest eigenvalue as minimum mode.  Broadly speaking, a minimum
mode following method maximizes along the direction of the minimum mode
and it minimzes in all other directions.  The optimization is
considered to be converged if the curvature along the minimum mode is
negative and if the norm of the gradient is smaller than a certain
threshold. As for the minimization, other force criteria are possible.

The minimum mode of the Hessian can be found by minimizing the curvature function
$c:\mathbb{R}^{3N} \mapsto \mathbb{R}$
\begin{align}
c(\vec{d}) &= \frac{\vec{d}^TH\vec{d}}{\vec{d}^T\vec{d}}\notag\\
           &\approx \frac{\Delta\vec{g}\cdot\Delta\vec{R}}{h^2},
\end{align}
where along with $h\ll1$ the following definitions were used:
$\Delta\vec{R}:=h\frac{\vec{d}}{|\vec{d}|}$ and
$\Delta\vec{g}:=\vec{\nabla}E(\vec{R}+\Delta\vec{R})-\vec{\nabla}E(\vec{R})$.
The vector $\vec{R}$ is the position at which the Hessian $H$ is evaluated at.
For the minimization of $c(\vec{d})$, we use the algorithm described in section~\ref{sec:findmin}
where the energy as objective function is replaced by $c(\vec{d})$. In
the pseudocode below, the here discussed minimization is done at line
6.
Under the constraint of normalization, the gradient
$\left.\vec{\nabla}c(\vec{d})\right|_{|\vec{d}|=1}$ is given by
\begin{align}
\left.\vec{\nabla}c(\vec{d})\right|_{|\vec{d}|=1} &=
2\left(H\vec{d}-c(\vec{d})\vec{d}\right)\notag\\
&\approx
2\left(\frac{\Delta\vec{g}}{h}-\left(\frac{\Delta\vec{g}\cdot\Delta\vec{R}}{h^3}\right)\Delta\vec{R}\right).\label{eq:curvgraddiff}
\end{align}
Blindly using the biomolecule preconditioner of
section~\ref{sec:bioprec} for the minimization of $c(\vec{d})$ would
mean that the gradient of Eq.~\ref{eq:curvgraddiff} was projected on
the bond vectors of
$\vec{d}$. Obviously, the bond vector as defined in
section~\ref{sec:bioprec} has no meaning for $\vec{d}$. Therefore,
Eq.~\ref{eq:curvgraddiff} instead is projected onto the bond vectors of
$\vec{R} + \Delta\vec{R}$.

At a stationary point, systems with free boundary conditions have six vanishing eigenvalues. The respective eigenvectors correspond to overall
translations and rotations.\cite{Wales2003} Instead of directly using
Eq.~\ref{eq:curvgraddiff} for the minimization of the curvature of
those systems, it is advantageous to remove the translations and
rotations from $\Delta\vec{R}$ and
$\left.\vec{\nabla}c(\vec{d})\right|_{|\vec{d}|=1}$ in
Eq.~\ref{eq:curvgraddiff}.\cite{Page1988,Mills1998,Wales2003}

The convergence criterion for the minimization of $c(\vec{d})$ has a
large influence on the total number of energy and force evaluations
needed to obtain convergence. It therefore must be chosen carefully.

The minimum mode is usually not computed at every iteration, but only
if one of the following conditions is fulfilled:
\begin{enumerate}
\item \label{item:first} at the first iteration of the optimization
\item \label{item:pathlength} if the integrated length of the optimization path
connecting the current point in coordinate space and the point at which
the minimum mode has been calculated the last time exceeds a given
threshold value $r_{\text{recomp}}$
\item \label{item:it} if the curvature along the minimum mode is
positive and the curvature has not been recomputed for at least $n_{\text{recomp}}$ iterations
\item \label{item:fnrm} if the curvature along the minimum mode is positive and the norm of the gradient falls below the convergence criterion
\item \label{item:tighten} at convergence (optional)
\end{enumerate}
In the pseudocode, these conditions are checked in line 5.  Among
these conditions, condition no.~\ref{item:pathlength} is, with respect
to the performance, the most important one. The number of energy and
gradient evaluations needed for converging to a saddle point can be
strongly reduced if a good value for $r_{\text{recomp}}$ is chosen.
Condition~\ref{item:it} and~\ref{item:fnrm} can be omitted for most
cases.  However, for some cases they can offer a slight reduction in
the number of energy and gradient evaluations. For example for the
alanine dipeptide system used in section~\ref{sec:bench}, these two
conditions offered a performance gain of almost 10\%.  Although
possible, we usually do not tune $n_{\text{recomp}}$, but typically use
$n_{\text{recomp}}=10$.  In our implementation,
condition~\ref{item:tighten} is optional. It can be used if very
accurate directions of the minimum mode at the saddle point are needed.
In this case, this last minimum mode computation can also be done at
a tighter convergence criterion. Further energy and gradient computations
are saved in our implementation by using the previously computed
minimum mode as the starting mode for a new curvature minimization.

As stated above, a saddle point is found by maximizing along the
minimum mode and minimizing in all other directions. This is done by
inverting the preconditioned gradient component that is parallel to the
minimum mode. This is shown at line 19 of the pseudocode below.  
For the case of biomolecules, the component of the bond-stretching
gradient that is parallel to the minimum mode is also inverted (line
13).
As already mentioned in section~\ref{sec:precondSigGrad}, the feedback
that adjusts the stepsize of $\vec{\nabla}E_{\perp}$ is slightly
different in case of the saddle finding method. Let
$\widehat{\vec{d}}_{\text{min}}$ be the normalized direction of the
minimum mode.  Then, in contrast to minimizations, the stepsize that is
used to scale $\vec{\nabla}E_{\perp}$ is not based on the angle between
the complete $\vec{\nabla}E$ and $\vec{\nabla}E^{\text{P}}$, but only on the angle between $\vec{\nabla}E-\left(\vec{\nabla}E\cdot\widehat{\vec{d}}_{\text{min}}\right)\widehat{\vec{d}}_{\text{min}}$ and
$\vec{\nabla}E^{\text{P}}-\left(\vec{\nabla}E^{\text{P}}\cdot\widehat{\vec{d}}_{\text{min}}\right)\widehat{\vec{d}}_{\text{min}}$.
These are the components that are responsible for the minimization in directions that are not the minimum mode direction.
Otherwise, the gradient feedback is absolutely identical to that
described in section~\ref{sec:precondSigGrad}.

A saddle point can be higher in energy than the configuration at which
the optimization is started at. Therefore, in contrast to a
minimization, it is not reasonable to discard the history, if the
energy increases. As a replacement for this safeguard, we restore to a
simple trust radius approach in which any atom must not be moved by more
than a predefined trust radius $r_{\text{trust}}$. A displacement
exceeding this trust radius is simply rescaled.  If the curvature is
positive and the norm of the gradient is below the convergence
criterion, we also rescale displacements that do not come from
bond-stretchings. The displacement is rescaled such that the
displacement of the atom that moved furthest, is finally given
by $r_{\text{trust}}$. This avoids arbitrarily small steps close to minima.

On very rare occasions, we could observe for some cluster systems that over
the course of several iterations a few atoms sometimes detach from the main
cluster. To avoid this problem, we identify the main fragment
and move all neighboring fragments towards the nearest atom of the main
fragment.

Below, the pseudocode for SQNS is given.
It contains 3 parameters explicitly. $\alpha'_{\text{start}}$ and
$\alpha'_{\text{s,start}}$ are initial step 
sizes that scale $\vec{\nabla} E_{\perp}$ and $\vec{\nabla}            
E_{\text{str}}$, respectively. $m'$ is the maximum length of the history
list from which the significant subspace is constructed. 

The path-length threshold $r_{\text{thresh}}$ that
determines the recomputation frequency of the
minimum mode is implicitly contained in line 5.
Lines 14 and 21 imply the trust radius $r_{\text{trust}}$.

Besides all the parameters that are needed for the minimizer
of section~\ref{sec:findmin}, line 6 additionally implies the finite
difference step size $h$ that is used to compute the curvature and its
gradient.

Line 18 implicitly contains the parameter $\epsilon$,     
which is described in section~\ref{sec:sigsub}
\begingroup
\allowdisplaybreaks
\begin{flalign*}
\text{ 1. }  &\alpha' \leftarrow \alpha'_{\text{start}};\text{ }
             \alpha'_s \leftarrow \alpha'_{\text{s,start}}; &\\
\text{ 2. }  &l \leftarrow 1; &\\
\text{ 3. }  &\text{Initialize }\vec{R}_l\text{ with coordinates;} &\\
\text{ 4. }  &\keyw{repeat} &\\
\text{ 5. }  &\hspace{\inda}\keyw{if }\text{recompute minimum mode }\keyw{then}  &\\
\text{ 6. }  &\hspace{\indb}\parbox[t]{\dimexpr\linewidth-\indbb}{
              \setlength{\hangindent}{0.5\inda}
              Use algorithm of section \ref{sec:findmin} and obtain a
              normalized minimum $\widehat{\vec{d}}_{\text{min}}$ of
              $c(\vec{d})$ at $\vec{R}_{l}$, use the previously computed minimum mode as input;}&\\
\text{ 7. }  &\hspace{\inda}\keyw{end if }  &\\
\text{ 8. }  &\hspace{\inda}\keyw{if}  \text{ optimizing biomolecule } \keyw{then} &\\
\text{ 9. }  &\hspace{\indb}\parbox[t]{\dimexpr\linewidth-\indbb}{
              \setlength{\hangindent}{0.5\inda}Compute
              $\vec{\nabla}E_{\text{str}}$ for $\vec{R}_{l}$, as
              outlined in section~\ref{sec:bioprec};} &&\\
\text{10. }  &\hspace{\indb}\parbox[t]{\dimexpr\linewidth-\indbb}{
              \setlength{\hangindent}{0.5\inda}Adjust $\alpha'_s$ based
          on the feedback described in section~\ref{sec:bioprec};} &&\\
\text{11. }  &\hspace{\indb}\vec{s} \leftarrow \alpha'_s
              \vec{\nabla}E_{\text{str}}; &\\
\text{12. }  &\hspace{\indb}\vec{g}_l \leftarrow \vec{\nabla}E(\vec{R}_{l}) - \vec{\nabla}E_{\text{str}}; &\\
\text{13. }  &\hspace{\indb} \vec{R}_{l} \leftarrow \vec{R}_{l}  - \vec{s}
               + 2\left(\vec{s}\cdot\widehat{\vec{d}}_{\text{min}}\right)\widehat{\vec{d}}_{\text{min}};&\\
\text{14. }  &\hspace{\indb}\parbox[t]{\dimexpr\linewidth-\indbb}{
              \setlength{\hangindent}{0.5\inda}
              Check for trust radius condition as described
              in section \ref{sec:findsad}. Rescale, if needed;}&\\
\text{15. }  &\hspace{\inda}\keyw{else}   &\\
\text{16. }  &\hspace{\indb}\vec{g}_l \leftarrow \vec{\nabla}E(\vec{R}_{l}); &\\
\text{17. }  &\hspace{\inda}\keyw{end if }  &\\
\text{18. }  &\hspace{\inda}\parbox[t]{\dimexpr\linewidth-\indaa}{
              \setlength{\hangindent}{0.5\inda}
              Based on the $\{\vec{g}_j,\vec{R}_j\}_{j\le l}$ in the
history list, compute the preconditioned
              gradient $\vec{\nabla}E^{\text{P}}$ as outlined
              in sections~\ref{sec:sigsub}~to~\ref{sec:bioprec};}&\\
\text{19. }  &\hspace{\inda} \vec{R}_{l+1} \leftarrow \vec{R}_{l}
              - \vec{\nabla}E^{\text{P}}
              + 2\left(\vec{\nabla}E^{\text{P}}\cdot\widehat{\vec{d}}_{\text{min}}\right)\widehat{\vec{d}}_{\text{min}};&\\
\text{20. }  &\hspace{\inda}\parbox[t]{\dimexpr\linewidth-\indaa}{
              \setlength{\hangindent}{0.5\inda}
              Check for trust radius condition and for fragmentation as described
              in section \ref{sec:findsad}. Rescale and fix fragmentation, if needed;}&\\
\text{21. }  &\hspace{\inda}\parbox[t]{\dimexpr\linewidth-\indaa}{
              \setlength{\hangindent}{0.5\inda}Adjust $\alpha'$ based on
              the gradient feedback described in section~\ref{sec:findsad};}&\\
\text{22. }  &\hspace{\inda}\keyw{if } l > m' \keyw{ then} &\\
\text{23. }  &\hspace{\indb}\parbox[t]{\dimexpr\linewidth-\indbb}{
              \setlength{\hangindent}{0.5\inda} Remove
              $\vec{R}_{l-m'}$ and $\vec{g}_{l-m'}$ from
              the history list;}&\\
\text{24. }  &\hspace{\inda}\keyw{end if }&\\
\text{25. }  &\hspace{\inda}l \leftarrow l + 1; &\\
\text{26. }  &\keyw{until convergence.} &\\
\end{flalign*}%
\endgroup

\section{Benchmarks and Comparisons}\label{sec:bench}
\begin{table*}
\caption{Benchmark results for minimizers. DFT test sets contain 100,
force field test sets contain 1000 distinct structures. SQNM runs
labeled with '(Bio)' indicate the usage of the preconditioner
for biomolecules described in section~\ref{sec:bioprec}.}
\begin{ruledtabular}
\begin{tabular}{cccrrrrrrrrrr}
&  & & & \multicolumn{4}{ c }{To Same Minimum} & &\multicolumn{4}{ c }{To
Arbitrary Minimum}  \\ \cline{5-8} \cline{10-13}
System  & Level of Theory & Method & $n_{\text{f}}$\footnotemark[1]
&$N$\footnotemark[2] & $\left<n_{\text{ef}}\right>$\footnotemark[3] &
$\left< r\right>$\footnotemark[4] &
$\left<n_{\text{woi}}\right>$\footnotemark[5]& &$N$\footnotemark[2] &
$\left<n_{\text{ef}}\right>$\footnotemark[3] & $\left<
r\right>$\footnotemark[4] & $\left<n_{\text{woi}}\right>$\footnotemark[5] \\
\hline
Alanine Dipeptide & DFT         & FIRE       & 0  & 93 & 454 & 14.01 & 7602& & 100 & 458 & 14.14 & 7662 \\
\tm{1}            & \tm{3}      & L-BFGS      & 2  & 93 & 185 & 23.41 & 3876& & 100 & 188 & 24.02 & 3941 \\
                  & \tm{4}      & SQNM (Bio) & 0 & 93 & 198 & 14.10 & 3711 & &100 & 207 & 14.29 & 3858 \\\cline{2-13}
                  & Force Field & FIRE       & 0 & 954 & 414 & 12.21 & -- & &1000 & 418& 12.35 & --\\
                  & \tm{5}      & L-BFGS      & 1 & 954 & 156 & 19.69 & -- & &999  & 158& 20.39 & --\\
                  &             & SQNM (Bio) & 0 & 954 & 188 & 12.38 & -- & &1000 & 192& 12.57 & --\\
\tm{2}            & \tm{6}      & SQNM       & 0 & 954 & 356 & 12.27 & -- & &1000 & 363& 12.49 & --\\\cline{1-13}
$\text{Si}_{20}$  & DFT         & FIRE       & 0& 46 & 139& 18.83 & 2458 & &100& 143 & 19.52& 2513\\
\tm{7}            &\tm{9}       & L-BFGS      & 30& 46 & 73 & 27.19 & 1677 & &70 & 74 & 31.26 & 1714\\
                  &\tm{10}      & SQNM       & 0 & 46 & 83 & 16.00 & 1740 & &100& 86 & 16.50 & 1784\\\cline{2-13}
                  & Force Field & FIRE       & 0 & 486 & 147 & 13.26 & -- & &1000 & 163& 15.32 & --\\
                  &\tm{11}      & L-BFGS      & 0 & 486 & 57  & 25.49 & -- & &1000 & 65& 30.44 & --\\
\tm{8}            &\tm{12}      & SQNM       & 0 & 486 & 72  & 10.82 & -- & &1000 & 81& 11.93 & --\\
\end{tabular}
\begin{tikzpicture}[overlay, remember picture]
  \draw[->] ([yshift=2ex]1.north) -- (2);
  \draw[->] ([yshift=2ex]3.north) -- (4);
  \draw[->] ([yshift=2ex]5.north) -- (6);
  \draw[->] ([yshift=2ex]7.north) -- (8);
  \draw[->] ([yshift=2ex]9.north) -- (10);
  \draw[->] ([yshift=2ex]11.north) -- (12);
\end{tikzpicture}
\end{ruledtabular}
\footnotetext[1]{Number of failed optimizations.}
\footnotetext[2]{Number of runs over which the averages are taken.}
\footnotetext[3]{Average number of energy and force calls (only successful runs).}
\footnotetext[4]{Average integrated path length of the optimization
trajectory in units of Bohr.}
\footnotetext[5]{Average number of wavefunction optimization iterations.}
\label{tab:benchmin}
\end{table*}

\subsection{Minimizers}
We compare the performance of the new SQNM method to the FIRE and
L-BFGS minimizers. We did not include the CG method in
this benchmark, because FIRE has previously been shown to be
significantly more efficient than CG.\cite{Bitzek2006}
Both FIRE and L-BFGS belong to the best optimizers
in their class.  With regard to the required number of energy and force
evaluation, L-BFGS is one of the best minimizers available for the
optimization of atomic systems. With respect to noise tolerance, the
same is true for FIRE. Although more efficient than FIRE, L-BFGS tends
to fail if there are inconsistent forces and energies due to
computational
noise.\cite{Bitzek2006} Such inconsistencies are unavoidable in
electronic structure calculations like for example DFT. 

For $\text{Si}_{20}$ clusters and the alanine dipeptide biomolecule,
benchmarks were performed both at DFT and force field level.
For L-BFGS we used the reference implementation of
Nocedal\cite{Nocedal1980,Liu1989} which is available from his website. We are not
aware of any references implementation of FIRE. However, FIRE is
straightforward to implement and thus we used our own code.
For the benchmarks of the minimizers at DFT level, all codes were coupled to the
BigDFT electronic structure code.\cite{Genovese2008,Mohr2014}
For the benchmarks at force field level, we used the Assisted Model
Building with Energy Refinement (AMBER) force field in the ff99SB
variant as implemented in
AMBER Tools\cite{Amber14} and the Lenosky Silicon
force field.\cite{Lenosky2000,Goedecker2002}

For alanine dipeptide and $\text{Si}_{20}$, we generated test sets by
running MD simulations at force field level. At force field level each
test set contains 1000 structures that were taken from the MD
trajectories. Subsets
containing 100 of these force field structures were used as benchmark
systems at DFT level.
For each method, we tuned the parameters at force field level for a
subset of 100 configurations. Identical parameters were used both at
force field and DFT level.  The $\text{Si}_{20}$ system was considered
to be converged as soon as the norm of the force fell below $1.0\e{-4}$
Hartree/Bohr. Even if far away from a stationary point, relatively
small forces can arise in alanine dipeptide. Therefore, a much tighter
convergence criterion of $1.0\e{-5}$ Hartree/Bohr had to be chosen for
alanine dipeptide.

Table~\ref{tab:benchmin} gives the results of these benchmarks. In
addition to the average number of energy and force calls
$\left<n_{\text{ef}}\right>$ we also give the average integrated path
length of the optimization path $\left<r\right>$. $\left<r\right>$ is
computed by summing all the distances between structures for which
consecutive energy and force evaluations were performed.

There is no guarantee that minimizations that are started at the same
configuration will converge to the identical minimum. Therefore,
Table~\ref{tab:benchmin} gives averages for both, the subset of runs
that all converged to identical minima and averages over all runs,
regardless of whether the final minima were identical, or not. Identical
configurations were identified by using the recently developed
s-overlap fingerprints.\cite{Sadeghi2013}

In all benchmarks, FIRE is clearly inferior to L-BFGS and SQNM. With
respect to the average number of energy and force evaluations, the
L-BFGS method is slightly more efficient than the new SQNM minimizer.
However, $\left<r\right>$ of L-BFGS is $1.6$ to $2.6$ times larger than
the corresponding values of the SQNM method.  On average, this means
that L-BFGS displaces the atoms more violently than SQNM. In DFT
calculations, the wavefunction of the previous optimization step can be
used as input wave function for the current iteration. Roughly
speaking, the less the positions of the atoms have changed, the better
this input guess usually is. Therefore, less wavefunction optimizations
are needed for convergence. To quantify this, the average number of
wavefunction optimization iterations   $\left<n_{\text{woi}}\right>$
needed for a minimization of the potential energy surface is
given in table~\ref{tab:benchmin}.  As a consequence of the smaller
displacements in the SQNM method, the L-BFGS and the SQNM method
roughly need the same number of wavefunction optimizations for
converging to a minimum of the potential energy surface.

The L-BFGS minimizer proved to be unreliable at DFT level. For example,
$30\%$ of all $Si_{20}$ minimizations failed to converge. In contrast
to this, all SQNM runs successfully converged to a minimum.

\subsection{Saddle Finding Methods}
\begin{table*}
\caption{Benchmark results for saddle finding methods. DFT test sets
contain 100, force field test sets contain 1000 distinct structures.
SQNS runs labeled with '(Bio)' indicate the usage of the preconditioner
for biomolecules described in section~\ref{sec:bioprec}.}
\begin{ruledtabular}
\begin{tabular}{cccrrrrrrrrrr}
&  & & & \multicolumn{2}{ c }{To Same Saddlepoint} && \multicolumn{2}{ c }{To
Arbitrary Saddlepoint}  \\ \cline{5-6} \cline{8-9}
System  & Level of Theory & Method & $n_{\text{f}}$\footnotemark[1] &$N$\footnotemark[2] & $\left<n_{\text{ef}}\right>$\footnotemark[3] &  &$N$\footnotemark[2] & $\left<n_{\text{ef}}\right>$\footnotemark[3] \\
\hline
Alanine Dipeptide & DFT         & SQNS        & 0 & --  & --  & & 100  & 510   \\\cline{2-9}
\tm{13}           & Force Field & DIMER       & 0 & 87 & 1324 & & 1000 & 3146  \\
                  & \tm{15}     & SQNS (Bio)  & 0 & 87 & 309  & & 1000 & 415  \\
\tm{14}           & \tm{16}     & SQNS        & 0 & 87 & 632  & & 1000 & 757  \\\cline{1-9}
$\text{Si}_{20}$  & DFT         & DIMER       & 0 & 8 & 234  & & 100 & 444  \\
\tm{17}            &\tm{19}       & SQNS      & 0 & 8 & 140  & & 100 & 237  \\\cline{2-9}
                  & Force Field & DIMER       & 0 & 20 & 264  & & 1000 & 622  \\
\tm{18}                  &\tm{20}      & SQNS & 0 & 20 & 189  & & 1000 & 368  \\
\end{tabular}
\begin{tikzpicture}[overlay, remember picture]
  \draw[->] ([yshift=2ex]13.north) -- (14);
  \draw[->] ([yshift=2ex]15.north) -- (16);
  \draw[->] ([yshift=2ex]17.north) -- (18);
  \draw[->] ([yshift=2ex]19.north) -- (19);
  \draw[->] ([yshift=2ex]20.north) -- (20);
\end{tikzpicture}
\end{ruledtabular}
\footnotetext[1]{Number of failed optimizations.}
\footnotetext[2]{Number of runs over which the averages are taken.}
\footnotetext[3]{Average number of energy and force calls (only successful runs).}
\label{tab:benchsad}
\end{table*}

The SQNS method was compared to an improved version of the dimer
method\cite{Henkelman1999} as described in Ref.~\citenum{Kaestner2008} and as implemented in the
EON code.\cite{Chill2014} In this improved version, the
L-BFGS\cite{Nocedal1980,Liu1989,Kaestner2008} algorithm is used for
the rotations and translation of the dimer. Furthermore, the rotational
force and the dimer energy are evaluated by means of a first order
forward finite difference of the
gradients.\cite{Heyden2005,Olsen2004,Kaestner2008} The same force
fields as for the minimization benchmarks were used. For the DFT
calculations, SQNS was coupled to the BigDFT code. The EON codes offers
an interface to VASP,\cite{Kresse1993,Kresse1994,Kresse1996,Kresse1996a,Kresse1999}
which consequently was used.

The same test sets as for the minimizer benchmarks were used. In
particular this means that the starting configurations are not close
to a saddle point and therefore these test sets are comparatively
difficult for saddle finding methods. Again, parameters were only tuned
for a subset of 100 configurations at force field level. With exception
to the finite difference step size that is needed to calculate the curvature and
its gradient, we used the same parameters at force filed and DFT level.
Because of noise, the finite difference step size must be chosen larger
at DFT level.
The same force norm convergence criteria as for the minimization
benchmarks were used. In all SQNS optimizations the minimum mode was
recalculated at convergence (condition 5 of section~\ref{sec:findsad}).

The test results are given in table~\ref{tab:benchsad}. In contrast to
the minimization benchmarks, we do not give averages for the number
wavefunction optimization iterations, because the two saddle finding
methods were coupled to two different electronic structure codes.
Therefore, the number of wavefunction optimizations is not comparable.

In particular in case of the $\text{Si}_{20}$ system, both methods
converged only seldom to the same saddle points and therefore the
statistical significance of the corresponding numbers given in
table~\ref{tab:benchsad} is limited. However, averages over large
sets could be made in the case of convergence to an arbitrary saddle
point.

In the cases we considered, the dimer method needed between $1.4$ and
$7.6$ times more energy and force evaluations than the new SQNS method.
In particular for alanine dipeptide, the SQNS approach was far superior
to the dimer method. Due to its inefficiency, it was impossible to
obtain a significant number of saddle points for alanine dipeptide at
DFT level when using the dimer method. For this reason, only benchmark
results for the SQNS method are given for alanine dipeptide at DFT
level.

\section{Conclusion}\label{sec:summary}
Optimizations of atomic structures belong to the most important routine
tasks in fields like computational physics, chemistry, or biology.
Although the energies and forces given by computationally demanding
methods like DFT are physically accurate, they are contaminated
by noise. This computational noise comes from underlying integration
grids and from self-consistency cycles that are stopped at non-vanishing
thresholds. The availability of
optimization methods that are not only efficient, but also
noise-tolerant is therefore of great importance. In this contribution we have presented a
technique to extract significant curvature information from noisy
potential energy surfaces. We have used this technique to create a
stabilized quasi-Newton minimization (SQNM) and a stabilized
quasi-Newton saddle finding (SQNS) algorithm. SQNM and SQNS were
demonstrated to be superior to existing efficient and well established
methods.

Until now, the SQNM and the SQNS optimizers have been used over a period of
several months within our group. During this time they have performed
thousands of optimizations without failure at the DFT level. Because of their
robustness with respect to computational noise and due to their
efficiency, they have replaced the default optimizers that have
previously been used in Minima
Hopping\cite{Goedecker2004,Goedecker2005} and Minima Hopping Guided
Path Search\cite{Schaefer2014} runs.

Implementations of the minimizer and the saddle search method are made
available via the BigDFT electronic structure package. The code is
distributed under the GNU General Public License and can be downloaded
free of charge from the BigDFT website.\cite{BigdftWeb}

\begin{acknowledgments}
We thank the Indo-Swiss Research grant for financial support. Computer
time was provided by the Swiss National Supercomputing
Centre (CSCS) under project ID s499.

\end{acknowledgments}



\providecommand{\noopsort}[1]{}\providecommand{\singleletter}[1]{#1}%
\end{document}